\documentclass[11pt,twoside]{article}
\usepackage{./asp2014}

\aspSuppressVolSlug
\resetcounters

\begin{document}

\title{Stellar Emission as a Source of Flux Bias in Debris Disks}
\author{Jacob Aaron White$^{1}$, Jason Aufdenberg$^{2}$, Aaron C. Boley$^{3}$, Peter Hauschildt$^{4}$, A. Meredith Hughes$^{5}$, Brenda Matthews$^{6}$, Attila Mo\'or$^{1}$, David J. Wilner$^{7}$}

\affil{$^{1}$Konkoly Observatory, Research Centre for Astronomy and Earth Sciences, Hungarian Academy of Sciences, Konkoly-Thege Mikl\'os \'ut 15-17, 1121 Budapest, Hungary, {\rm jacob.white@csfk.mta.hu}}

\affil{$^{2}$Physical Sciences Department, Embry-Riddle Aeronautical University, 600 S Clyde Morris Blvd., Daytona Beach, FL 32114, USA}

\affil{$^{3}$Department of Physics and Astronomy, University of British Columbia, 6224 Agricultural Rd., Vancouver, BC V6T 1T7, Canada}

\affil{$^{4}$Hamburger Sternwarte, Gojenbergsweg 112, 21029 Hamburg, Germany}

\affil{$^{5}$Department of Astronomy, Van Vleck Observatory, Wesleyan University, 96 Foss Hill Dr., Middletown, CT 06459, USA}

\affil{$^{6}$Herzberg Institute, National Research Council of Canada, 5071 W. Saanich Road, Victoria, BC V9E 2E7, Canada}

\affil{$^{7}$Harvard-Smithsonian Center for Astrophysics, 60 Garden Street, Cambridge, MA 02138, USA}

\paperauthor{Jacob Aaron White}{jacob.white@csfk.mta.hu}{0000-0001-8445-0444}{Konkoly Observatory}{}{Budapest}{}{1121}{Hungray}

\begin{abstract}

Our understanding of stellar atmospheres and our ability to infer architectures of extrasolar planetary systems rely on understanding the emission of stars at sub-millimeter to centimeter wavelengths. In this chapter we describe how unconstrained stellar emission can interfere with the accurate characterization of circumstellar debris. The ngVLA is the only facility with the sensitivity that allows for the observations of a broad range of stellar spectral types in a feasible amount of time. The observations will enable the building and testing of accurate models of stellar emission, which in turn are required for evaluating both the occurrence and abundance of debris over the proposed wavelength range of the ngVLA. 
 
\end{abstract}

\section{Introduction}

The next-generation Very Large Array (ngVLA), with its proposed technical specifications, will extend the search and study of debris disks to much further distances than is currently possible. The high sensitivity over the proposed wavelength range will also allow for the study of a previously difficult (and sometimes impossible) to explore size range of disk particles. For a given debris system, if the disk emission is unable to be separated spatially from its host star, an accurate model of the stellar emission is required in order to constrain the flux contributions of the star and the disk. This is particularly true for warm/hot debris systems such as exo-zodiacal dust and exo-asteroid belts.

Emission from stars over the proposed wavelength range of the ngVLA is nontrivial and generally not well-constrained \citep{cranmer}, with the Sun being the most thoroughly studied star in this regime \citep[see][for other G-type stars]{liseau15}. In the atmosphere of the ``quiet'' Sun, the submm/cm continuum radiation is due primarily to free-free emission \citep{dulk, loukitcheva}. Quiet Sun models predict a 1 mm brightness temperature ($\rm T_{B}$) of $\sim 4700$ K, or $\sim 80\%$ of the Sun's photosphere $\rm T_{B}$ \citep{dulk}. The ``active'' Sun, with strong magnetic fields, is difficult to model because there are many contributing emission mechanisms \citep[e.g., umbral oscillations and explosive events;][]{wang, wedemeyer16}. The $\rm T_{B}$ spectrum of the Sun varies significantly, with a minimum in the far-infrared/submm. This is followed by a pronounced increase in flux at mm wavelengths and very high $\rm T_{B}$ at cm wavelengths (see Fig.\,1). 

The Sun cannot be used as a template for all stars, however, due to the differences in stellar atmospheres and magnetic activity. This is shown in Fig.\,1 with observations of A-type stars Sirius A, Altair, and Fomalhaut. In contrast to the Sun, the Sirius A data and corresponding PHOENIX models show that the brightness temperature continues to decrease with increasing wavelength over the observed range \citep{white_mesas}. Together, the Sun and Sirius A show that when considering the flux from a star, it can not be assumed that the submm/cm $\rm T_{B}$ is well represented by the photosphere's $\rm T_{B}$. It may also not be reasonable to extend the far-infrared brightness temperature to longer wavelengths, as is often done in the literature. Further still, it can not be assumed that a given spectral type, for example an A-type star, will have the same general $\rm T_{B}$ spectrum as another A-type star. This again is seen in Fig.\,1 when comparing Altair, Fomalhaut, and Sirius A and is discussed further in Sec.\,2. 

In this chapter, we explore how the ngVLA can be used to observe a large number of stars over the proposed wavelength range. These observations can be used to inform stellar atmosphere models (e.g., PHOENIX), which must be accurately known to study unresolved circumstellar debris.

\articlefigure{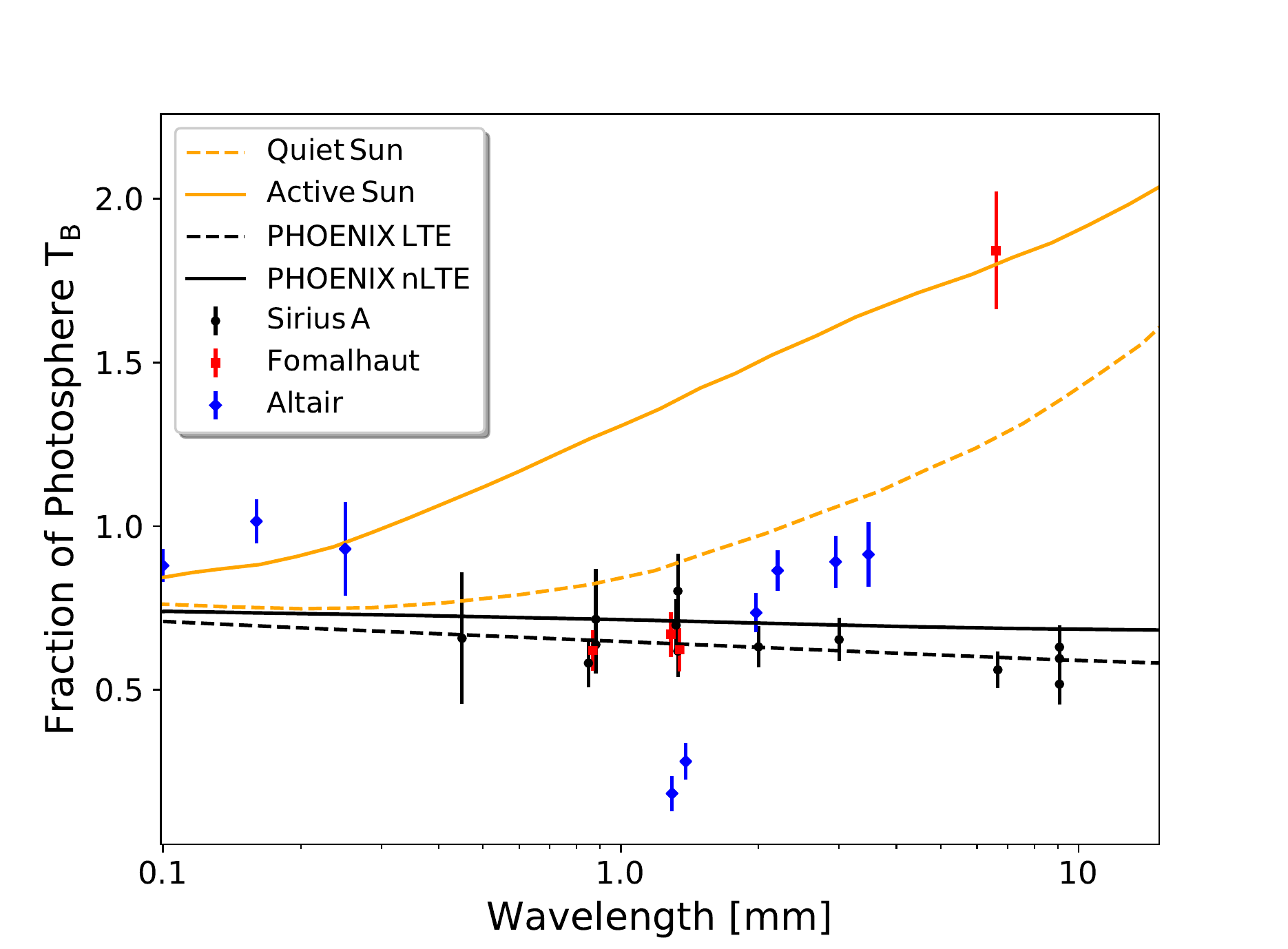}{fig:spec}{Temperature spectrum of the Sun, Sirius A, Fomalhaut, and Altair. The y-axis is the fractional brightness temperature ($\rm T_{B}$), or the observed $\rm T_{B}$ normalized with the photosphere $\rm T_{B}$ of a given star. This allows for different stellar types to be easily compared on the same plot. The two orange curves represent models of the Sun at maximum activity (solid line) and minimum activity (dashed line) from \citet{loukitcheva}. The black points are JCMT, SMA, ALMA, and VLA observations of Sirius A (White et al.~2018) and the two black curves are PHOENIX models of Sirius A's atmosphere with a non-LTE model (solid line) and a LTE model (dashed line). The blue diamonds are \textit{Herschel} and NOEMA observations of Altair (Thureau et al.~2014, White et al. in prep.) The red squares are ALMA and ATCA observations of Fomalhaut \citep{ricci, su16, white_fom, macgregor17}. }

\section{Implication for Debris Disks}

Poor characterizations of the radio flux from the host star in a debris system can non-trivially affect measurements of the occurrence and abundance of circumstellar debris. To see this, consider the following cases. Fomalhaut is an A3V star with a well-known extended debris ring at 140 au from the star and additional potential IR excess at much closer orbital distances \citep[e.g., an asteroid-like belt;][]{acke}. However, adopting different stellar models significantly changes the amount of this inferred excess \citep{su16, white_fom}. In particular, ALMA observations at 0.87 mm and 1.3 mm find a $\rm T_{B}$ of $\sim 5500$ K for the central emission. This is much lower than the optical photosphere temperature of 8650 K, and even lower than extrapolating far-IR emission. Confusingly, ATCA 7 mm observations find a $\rm T_{B}$ of $> 16000$ K \citep{ricci}, nearly $\sim 200\%$ of $\rm T_{B}$(phot), potentially showing behavior similar to the Sun (see Fig.\,1). No conclusions on a potential unresolved asteroid belt can be made until the emission of the Fomalhaut star is known and subtracted from the observed flux \citep{white_mesas, white_dis}.

HD 141569 is a 5 Myr-old system that contains a B9.5V star, an extended gas and dust structure, and two M dwarf companions at large separations. In 2014, unresolved 9 mm VLA observations of HD 141569 detected a higher than expected disk flux \citep{macgregor16}. This led to an inferred spectral index that was much shallower than average for disks, with interesting implications for the grain size distribution and dust evolution models. High resolution VLA follow-up observations in 2016, however, did not detect the expected dust structure. Instead, these observations found that the recovered emission was consistent with a point source centered on the star \citep{white_141569}. The total emission between the two semesters was also different by a factor of a few, suggesting that stellar emission, which appears to be variable, is affecting the 9 mm flux density and thus the interpretation of the dust in the system.

Another case where uncharacterized stellar emission has confounded debris disk studies is with Proxima Centauri. ALMA observations of Proxima found an unresolved excess over the ``expected'' stellar emission at 1.3 mm, in two different antenna configurations \citep{anglada}. This excess was initially interpreted as emission from multiple debris components in the system. Follow-up analysis of the ALMA data found instead that the observations caught a very large, short lived flare \citep{macgregor18}. While flaring is common in M-type stars (such as Proxima), a radio flare of $\sim1000\times$ the quiescent emission was not expected as it is nearly $10\times$ stronger than the strongest Solar flare ever detected \citep{krucker}. While Proxima is indeed an M-type star, with very different emission mechanisms than the A-type stars discussed above, the ALMA observations of Proxima highlight the importance for proper characterization of stellar emission in general.

\section{Uniqueness to ngVLA Capabilities}

\articlefigure{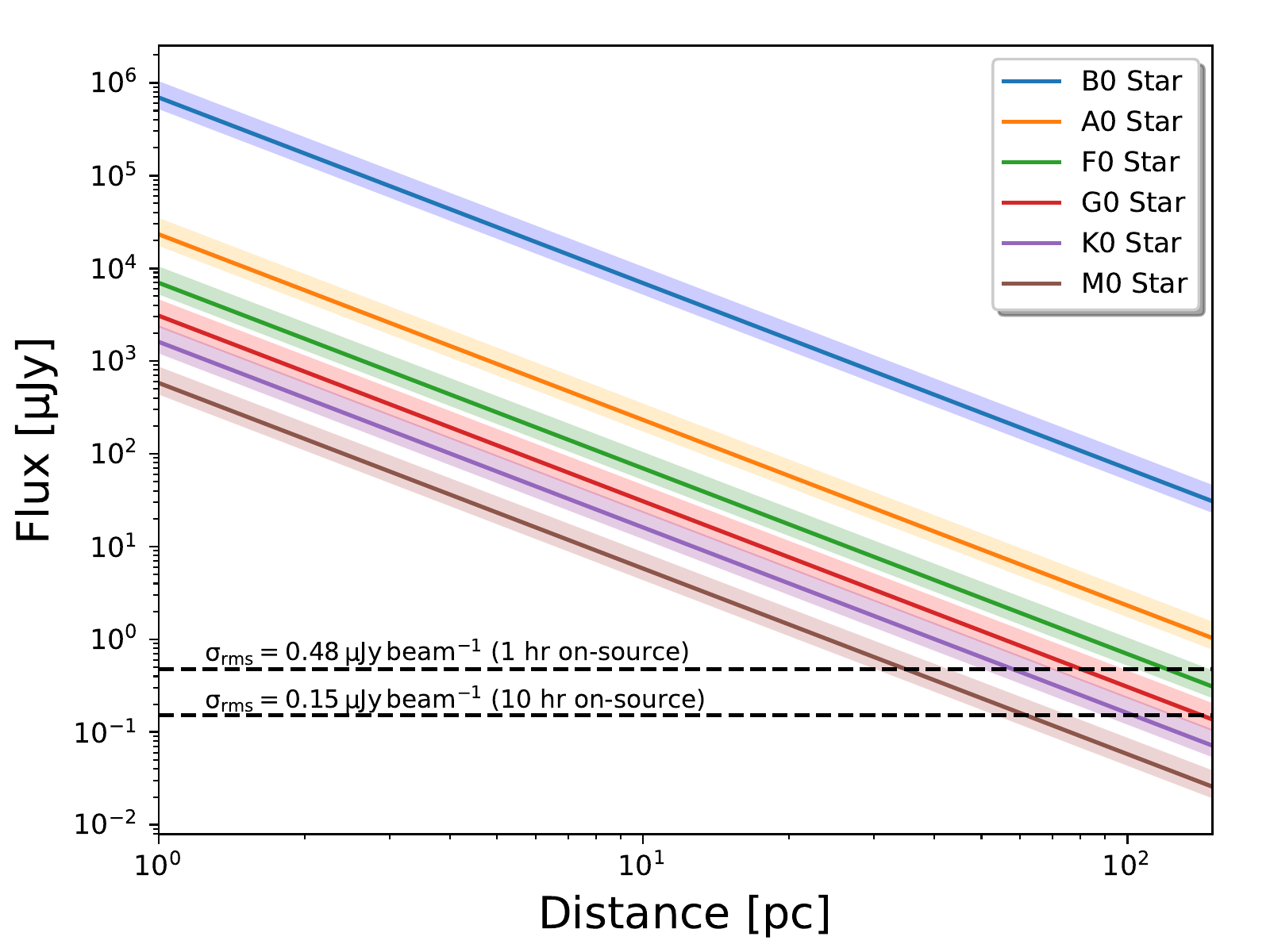}{fig:flux}{Estimated ngVLA Band 6 (93 GHz) flux of a given spectral type of star as a function of distance. The shaded region for each spectral type is the estimated stellar flux if the emission is best represented by a brightness temperature ranging from 75\% to 150\% of the photosphere temperature. The lower mass stars will likely have significant non-thermal emission due to, e.g., coronal activity and stellar magnetic fields (see orange curves in Fig\,1. This non-thermal emission can lead to significantly higher emission than shown here, making these estimates lower level limits of the expected flux. The estimated $\sigma_{\rm rms}$ from 1 hr and 10 hr on-source is given by the two horizontal dashed lines \citep{selina18}.  }

The ngVLA will be the only facility capable of detecting the radio emission from a wide range of debris-poor stars over the proposed wavelength range. While the ngVLA will indeed be able to resolve some nearby stars (the 0.7 mas maximum resolution corresponds to $\sim1~\rm D_{\odot}$ at a distance of $\sim15$pc), the success of this study is not dependent on resolving capabilities. Instead, this study relies on the high sensitivity provided the collecting area of the proposed 214 antennas. 

Current limitations on building a complete catalog of the radio emissions of stars is largely based on the required on-source observing times. For example, ALMA observations of Sirius A highlight this issue. Sirius A is the closest A-type star to Sun at 2.6 pc and therefore is the easiest target to observe. ALMA 3 mm observations of Sirius A only required $\sim1$ min of on-source observing time to achieve a SNR of 15 (White et al.~in prep). To observe a Sirius-like star that is 15 pc away, it would take over 18 hr of on-source observing time. This makes observing anything but the closest and brightest debris-poor stars very difficult to do with current facilities.

In order to build a full catalog of the radio emission of stars with the ngVLA, a targeted survey approach can be adopted. The closest and brightest stars will be observed first with broad spectral coverage. With a proposed 93 GHz continumm sensitivity of $\sigma_{\rm rms} = 0.48~\mu\rm Jy~beam^{-1}$ \citep{selina18}, these sources will only require on the order of minutes on-source time to get a useful SNR (e.g., >10). The observations will be used to inform stellar atmosphere modeling codes \cite[e.g., PHOENIX;][]{hauschildt}. The feasibility of this has already been demonstrated with Sirius A \cite{white_mesas}. The survey will then begin extending out to more distant sources, targeting a range of stellar types and properties (e.g., rapid rotators, metal rich/poor, etc.). Templates of stellar emission profiles will be made readily available so that they can be utilized in properly characterizing unresolved debris structures observed with the ngVLA.

\textit{IRAS} and \textit{Spitzer} surveys at 70 and 100 $\mu$m found that A-type stars are the most likely to host a detectable debris component \citep{su06, thureau}. This high occurrence rate of debris makes A-type stars common targets in studies that seek to characterize debris (which in turn requires an accurate model of the star's spectrum). As can be seen in Fig.\,2, nearly all A-types within 150 pc will be detectable in 1 hr of observing time if their emission is characterized by $>75\%$ of their photosphere $\rm T_{B}$. Within 10 hr of observing time, most stars of a size and temperature greater than that of typical early G-type stars will be detectable. With current stellar number density estimates within 150 pc \citep[e.g.,][]{bovy}, this allows for hundreds of stars to be observable with the ngVLA in a reasonable amount of observing time.

The scope of this project also offers a unique opportunity for adding many new calibrators to the ngVLA's observational setup. Well-constrained stellar spectral profiles of nearby stars that exhibit no detectable variability will increase the on-sky dispersion of available flux and phase calibrators. A larger number of available calibrators can help reduce the overhead time in a given observation. If these stars were indeed used as calibrators, then it would increase the amount of data available to accomplish the overall goals of this proposed project.

\section{Synergies at Other Wavelengths}

A broad spectral profile is needed in order to fully characterize the structure of a given stellar atmosphere. Similarly, a broad spectral profile of a debris disk is needed in order to accurately constrain disk properties such as morphology, mass, and grain size distribution. Current observational facilities such as ALMA, NOEMA, and SMA can probe the submm-mm wavelength emission from debris disks. Future facilities such as the SKA may be able to probe the radio emission from debris disks at even longer wavelengths than the ngVLA. The ``gap'' between these two wavelength regimes is also under-explored in regards to stars and debris due to current sensitivity constraints, making the ngVLA a key tool in to understand the physics that occur in this regime. Only by combing data from all these facilities can we observe the full spectral profile of a star, build and test models of its emission, and use these models to studying unresolved debris disks.

\section{Summary}

The ngVLA will provide unprecedented sensitivity over the proposed wavelength range, creating the opportunity to observe debris disks at much further distances than currently possible. In unresolved disks, an accurate determination of the flux contribution from the disk and the star requires a well tested model of stellar emission. Current observational facilities are only able to observe a few nearby debris-poor stars, making it difficult to test models of stellar atmospheric emission. The ngVLA will have the sensitivity required to observe significantly more stars, including essentially all A-type stars within 150 pc. These observations will be absolutely essential in order to accurately constrain the frequency and abundance of circumstellar debris.




\end{document}